# Superconductivity in Yttrium Iron Oxyarsenide System

**S V Chong, T Mochiji and K Kadowaki**

Institute of Materials Science and Graduate School of Pure & Applied Sciences, University of Tsukuba, 1-1-1, Tennodai, Tsukuba, Ibaraki 305-8573, Japan

E-mail: s_chong@ims.tsukuba.ac.jp

**Abstract**. Iron-based oxypnictides substituted with yttrium have been prepared via a conventional solid state reaction. The product after first 50 hours of reaction showed diamagnetic-like transition at around 10 K but was not superconducting, while additional 72 hours of high temperature heat treatment was required to yield superconducting sample which was doped with fluoride. Temperature dependence of the susceptibility shows both screening and Meissner effect at around 10 K, while resistance as a function of temperature displayed a drop at around the same temperature.

## 1. Introduction

Superconductivity in lanthanide (*Ln*) substituted iron oxypnictides (*Ln*FeAsO) doped with electrons or holes have been a fascinating research topic ever since the discovery of $LaFeAsO_{1-x}F_x$ having a superconducting transition temperature ($T_c$) of 26 K [1]. To date, the highest $T_c$ being reported were found in electron doped $Gd_{0.8}Th_{0.2}FeAsO$ prepared under ambient pressure and $SmFeAsO_{0.9}F_{0.1}$ prepared at high pressure both having $T_c$ of around 55 K [2-3]. It is known that reducing the lattice parameters of the tetragonal crystal structure, which can be viewed as increasing the internal pressure, plays an important role in promoting higher $T_c$ [4]. This has been achieved by the substitution of smaller lanthanides: these heavier lanthanides all exhibit higher $T_c$'s compared to that of lanthanum substituted in the order of La < Ce < Pr, Sm, Nd, etc [5]. Another physical means of achieving this have been in preparing the oxypnictides under high pressures, which have been of great success in preparing fluoride and non-fluoride doped superconductors [2,5,6], and including one of the first reported $SmFeAsO_{1-x}F_x$ single crystals [7].

Yttrium being a smaller non 4f element just above lanthanum in the periodic table and an important constituent in cuprate-based high temperature superconductors (HTS) has been widely predicted to be a good substitution candidate in oxypnictide compounds [8,9]. Nekrasov *et al.* have preformed *ab initio* calculations on the electronic structure of several F-doped iron-based superconductors, in which the authors also hypothesized that F-doped YFeAsO should exhibit superconductivity [9]. To date no YFeAsO quaternary compound has been reported in both the pure and doped form. Herein, we show that we were able to make YFeAsO to be superconducting successfully.

## 2. Experimental

Polycrystalline $YFeAsO_{1-x}F_x$ samples were prepared via conventional solid state reaction. Stoichiometric amount of YAs, FeAs, Fe, $Fe_2O_3$, and $YF_3$ (for doping) were finely grinded and thoroughly mixed before being pressed into pellets. YAs was prepared by reacting yttrium pieces and arsenic chips sequentially at 500 °C for 10 hours, 700 °C for 5 hours, and finally at 900 °C for 5 hours. FeAs was also prepared by reacting pressed pellet of finely ground 1:1 molar ratio mixed powders of iron and arsenic at 700 °C for 10 hours. The undoped and fluoride doped (x = 0.1) yttrium oxy iron-arsenide pellets were firstly reacted sequentially at 700 °C for 5 hours, then at 900 °C for 5 hours before being raised to 1150 °C for 50 hours. After checking for superconductivity and structural characteristic, the pellets were re-grounded and pressed into pellets and reacted furthermore at 1200 °C for ~72 hours. All reactions including the preparation of metal arsenides were carried out in vacuum sealed quartz tubes. The whole sample preparation excluding the annealing of samples were carried out in a glove-box filled with pure argon gas.

## 3. Results and Discussion

Temperature dependence of the mass susceptibility ($\chi_{mass}$) measurements (*M-T*) at 10 Oe applied field on the samples heated firstly at 1150 °C display diamagnetic-like transition at around 10 K in both the 0.1 F-doped (figure 1) and undoped (not shown) YFeAsO. A broad peak perhaps due to ferromagnetic transition was also observed at around 62 K in both samples indicating the presence of mixed multiple phases. This is supported by magnetization versus applied field (*M-H*) measurement at 5 K where both coercivity and remanence were most obvious compared to those at 110 K. Indeed x-ray powder diffraction shows the presence of large amount YAs, $Y_2O_3$, and FeAs. Since $Y_2O_3$ was not one of the starting materials, the formation of this metal oxide indicates that the occurred chemical reaction was improper or insufficient owing to an inadequate temperature or annealing duration.

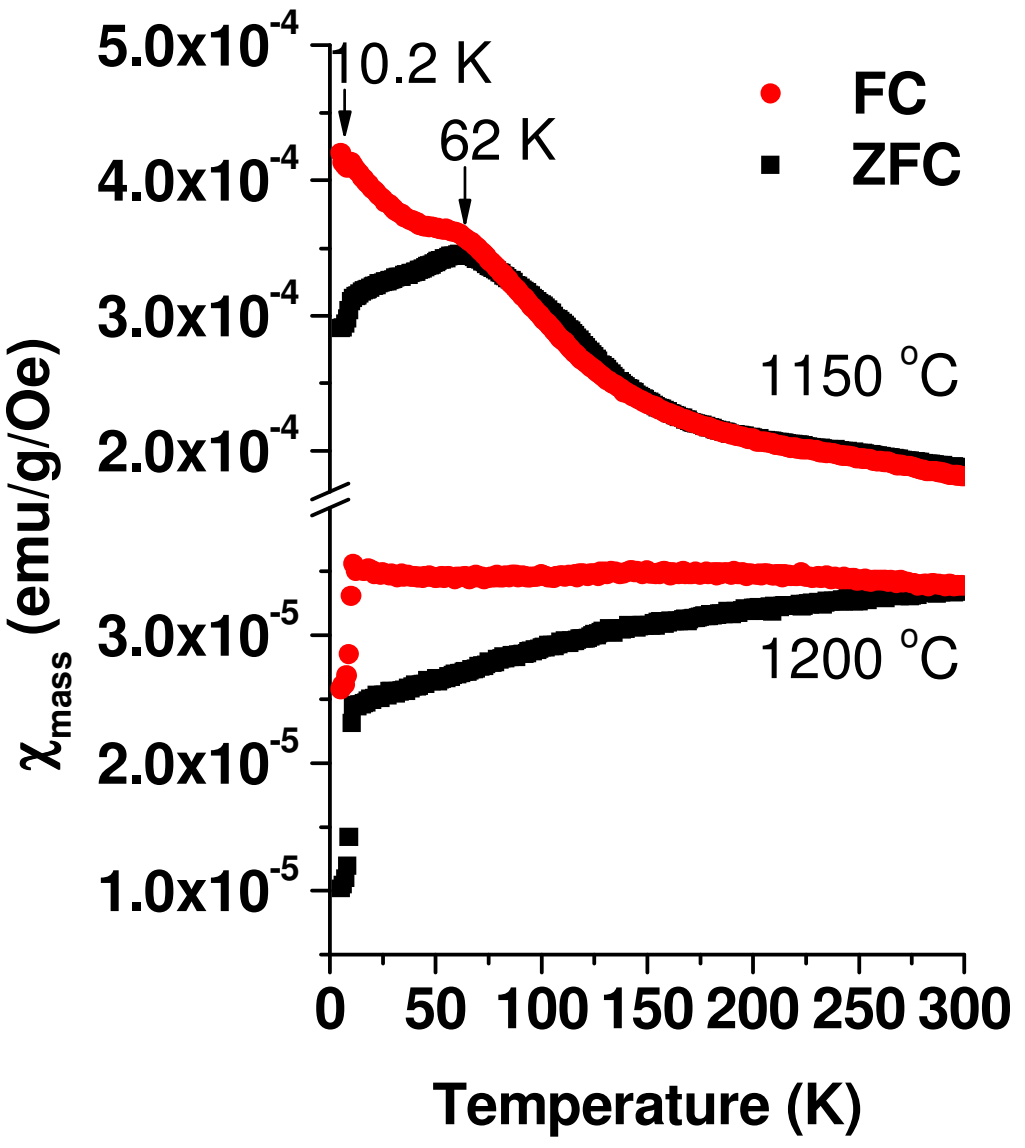

**Figure 1.** Temperature dependent mass susceptibility of the 0.1 F-doped sample reacted at different temperatures.

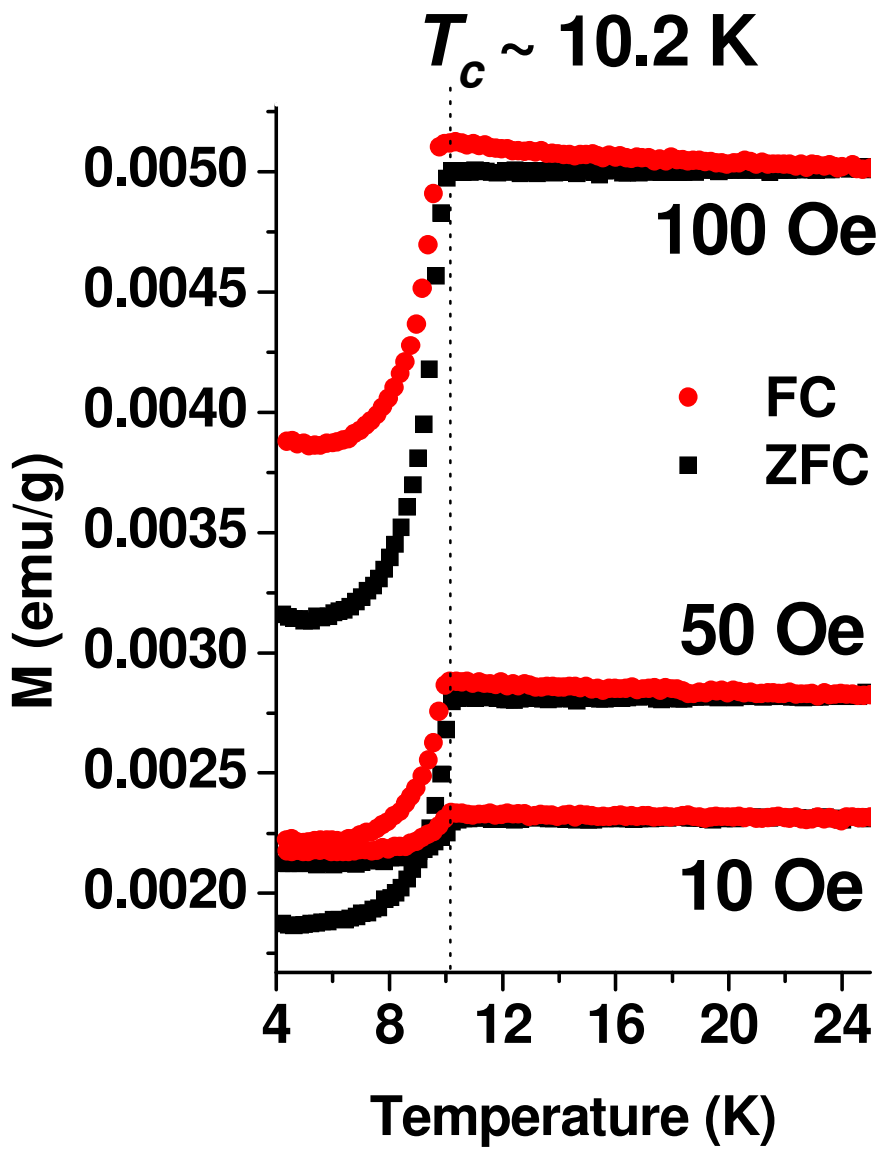

**Figure 2.** *M-T* measurements at different fields for the 0.1 F-doped sample.

After reacting the samples for a second time at 1200 °C, the zero field cooled (ZFC) and field cooled (FC) curves of the doped sample show both magnet screening and Meissner transition at 10.2 K as shown in the lower curves of figure 1. More detailed *M-T* scans at different fields up to 100 Oe confirmed the presence *of* these transitions (figure 2). Furthermore, the ferromagnetic transition peak at around 65 K was completely attenuated. Temperature dependent resistance measurement (*R-T*) displays metallic behavior below room temperature and the existence of two transition temperatures (figure 3). A small transition was observed first at 10.2 K which was followed by a larger and sharp transition at 9.3 K. A residual resistance down to 4 K was also observed in the sample, which indicates substantial amounts of non-superconducting phases in the sample. In fact it was difficult to identify the Bragg reflections related to the tetragonal *Ln*FeAsO structure as other reflection lines were superimposing on these. Comparing the most intense Bragg reflections in the doped and undoped samples, the reflection of the former was observed to be well resolved into two reflections with d-spacing values of 2.907 and 2.898 Å whereas that of the undoped sample has a broad peak centered at 2.907 Å as shown in figure 4. The lower d-pacing Bragg reflection is most likely associated to yttrium oxide and YAs as it was also observed in the both the doped and undoped sample reacted to 1150 °C. We have therefore tentatively assigned the reflection at 2.907 Å to that of YFeAsO since this is within the range of the (102) reflections of LaFeAsO at 2.965 Å [1] and 2.831 Å in the newly discovered TbFeAsO superconductor being the smallest lanthanide constituent reported in this superconductor family [10]. Interestingly, the undoped YFeAsO sample heated to 1200 °C also display diamagnetic-like transition at around 10 K in the temperature dependent magnetization curves, but no $T_c$ was observed in the resistance measurement as a function of temperature.

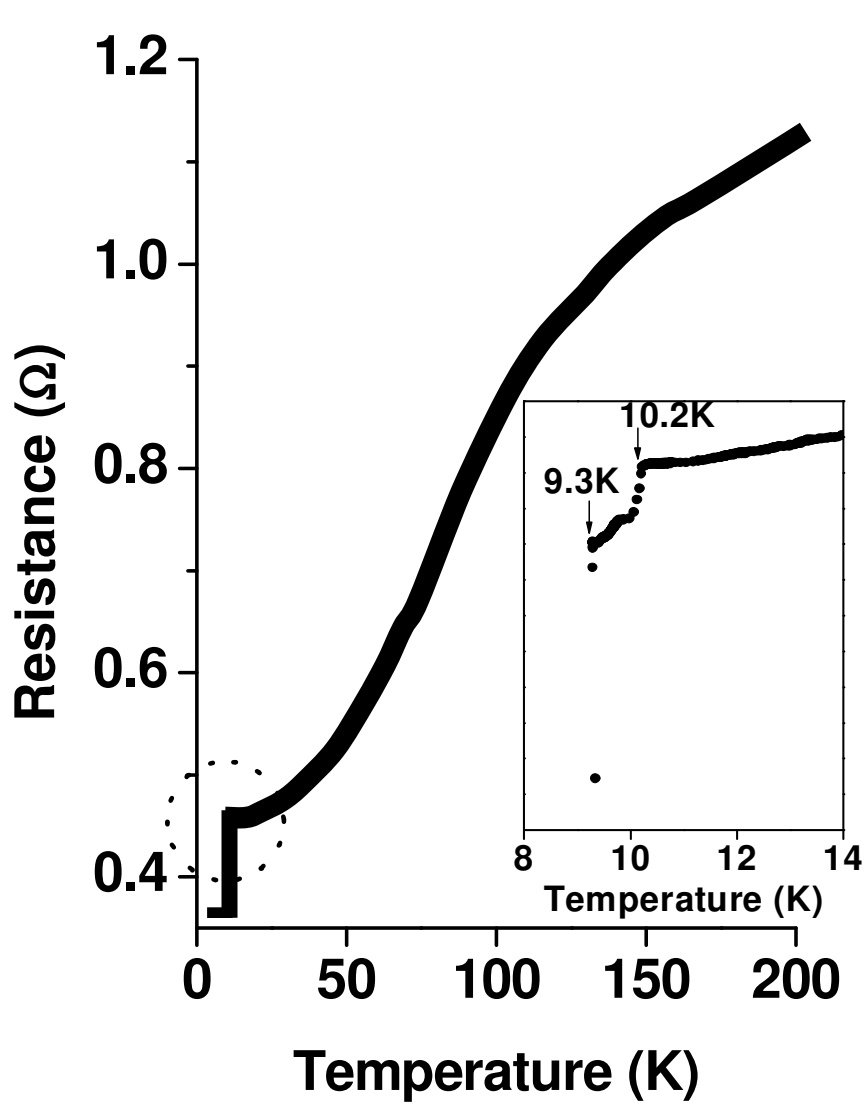

**Figure 3.** Temperature dependence of the resistance measurement of a 0.1 F-doped sample. The inset shows the enlargement of the circled part of the *R*-*T* curve.

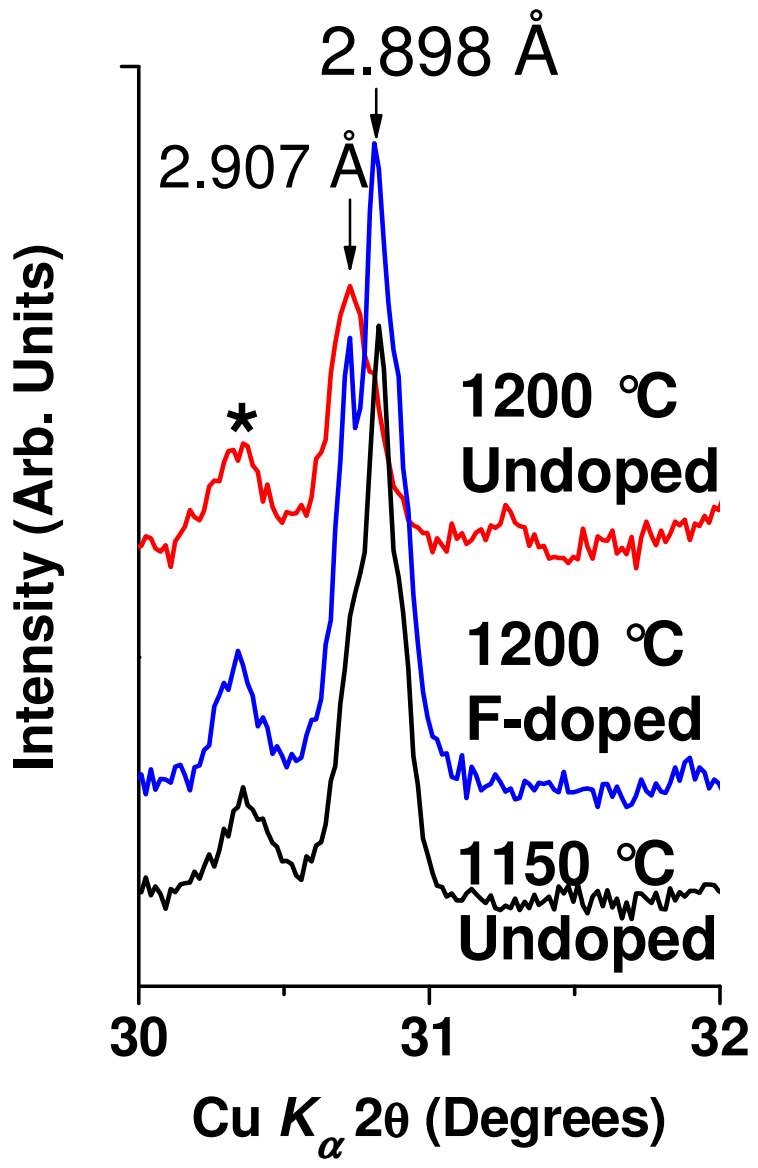

**Figure 4.** Powder x-ray diffraction patterns comparing the doped and undoped samples prepared at different temperatures. The asterisk indicates an impurity phase.

In summary, although completely pure phase of YFeAsO quaternary compounds could not be prepared, we reported here the observation of superconductivity in $YFeAsO_{0.9}F_{0.1}$ with an onset $T_c$ of 10.2 K.